# Through Global Monitoring to School of the Future: Smartphone as a Laboratory in Pocket of Each Student

Vladimir Shatalov[1], Victor Martynyuk[2], Maxim Saveliev[1]

[1] *National Technical University of Ukraine "Kiev Polytechnic Institute", Slavutych branch, 07101, Ukraine; E-mail: vladishat@gmail.com*
[2] *Taras Shevchenko National University of Kyiv, 01601, Ukraine*

**Abstract:** *The idea to unite smartphones used as personal environmental sensors and health indicators into a scalable network for data collection and processing by the internet-cloud is proposed. Access to the sensors, which are available in every smartphone, will provide the appropriate software. Such a monitoring at the global level would reveal the impact of the electromagnetic radiation, environmental pollution and weather factors on human health. Participation of students in these measurements increases their educational and social activities.*

## Introduction

A number existing problems in medicine, ecology, life science as well as in human being is related to the global factors as the root cause of the adverse effects observed. For example, these are the following problems.

*A*. Hypersensitivity of some people to weather factors [1]. That is why weather forecasts are often accompanied by warnings of a medical nature.

*B*. It is believed that the variations of solar activity are responsible for the increased risk of cardiovascular and other chronic diseases [2].

*C*. Some scientists believe that "space weather" affects antisocial moods of people, and the myths and legends of different countries link wars and other unhappiness with cosmic phenomena since ancient times [3].

*D*. Some elderly people, as well as persons requiring emergency medical care, with severe injury or persons wounded in the battlefield – they all need continuous monitoring of the major indicators of the cardiovascular system.

*F*. Pupils need a supervision of their activity and health status.

All of these cases require feedback in order to get recommendations of experts, medical advices or background information.

*The aim of the paper*

To solve the abovementioned problems we purpose a technology of the global monitoring of the human cardiovascular and environmental parameters (radiation, electromagnetic fields, and atmospheric) using smart phones and other gadgets that people have with post processing of data in a "cloud". This global monitoring will help scientists to separate the random noise from the correlated fluctuations and allow ones to draw conclusions about the reliability and causal relationship of the observed phenomena. Those conclusions may serve as a basis for recommendations on the feedback.

The same approach has been proposed recently [4] for observing cosmic rays at ultra-high energy. The authors assume to use existing network of smartphones as a ground detector array. In the present paper a similar approach is extended on to the scientific, medical, and educational purposes when every smartphone may get an informative feedback from the network.

### How it will work

*Smartphone*

Almost every smartphone is equipped with various sensors – namely, the pressure, humidity, light, linear acceleration, temperature, proximity, gravity sensors, and with an accelerometer, gyroscope, inclinometer, barometer, magnetometer and other micro-electro-mechanical systems (MEMS). The sensor's data may be read by software and processed based on the physical laws that describe the observed phenomena.

*Internet*

The data obtained in the different smartphones are transmitted through the Internet in the "cloud" infrastructure. In this way the data can be processed and combined into a common database. In addition, all participants will have access to these data. The data can be displayed at the map to create a picture of the ecological state, both for the close to the client areas and for the other regions involved in the project. The measurement experiments carried out in such a way can be used as a basis for collaboration and scientific research not limited to one institution, city or country.

### Global monitoring

People with the appropriate smartphones join the project for free (anonymously if needed). Every participant has to download special software to his smart-phone. This application monitors permanently the human body status along to the environmental parameters and sends the data to the "cloud" in Internet. The cloud processes the information and gives out recommendations or solutions. The client gets back the results of the monitoring which are concerning their own health, as well as the information on some of the global processes affecting human health and the recommendations of experts on the cases.

There are several environmental parameters available on smartphones right nowadays by the help of appropriate software, for example:
− Radioactivity Counter is using the camera sensor to detect radiation, like a Geiger-Mueller counter, of course with a smaller area.

- EMF Meter detects an electromagnetic field (EMF) emitting entity and/or object by using mobile phone's hardware as an EMF Sensor.
- Sound Meter measures loudness or sound pressure in decibels.
- Temperature and Pressure Meters.

The existing software may measure the following human health parameters.
- Cardiograph uses the device's built-in camera or dedicated sensor to calculate your heart's rhythm. The same approach is using in the professional medical equipment.
- Real Blood Pressure Calculator calculates people's blood pressure in two steps with approximately ±10%.

In such a way we get a large volume of data distributed across the globe that will allow us to carry out the identification of the correlated fluctuations of environmental parameters. Simultaneous monitoring of the health parameters of a large number of people will allow us to clear up the origin of the observable fluctuations in the health status of people.

**New advances in education**
*Extension classes outside the classroom*

All the students of the current generation dream to work in electronic school. Even now classes in some schools use computers, multimedia projectors, electronic sensor boards and more. As we believe in the school of the future information technologies will be extended beyond the classroom aiming to get classes of the nature itself.

Almost every student has in his pocket a smartphone. Actually, as it was mentioned in [5], this device can be easily transformed by special software into a small pocket laboratory that enables students to carry out some laboratory works and scientific observations. One can measure physical parameters of environment such as the background radiation, level of electromagnetic pollution, level of lighting, noise, and so on. Moreover, using the Global Positioning System (GPS) all these parameters may be linked to the place of observation. At the same time these smartphones can be used to measure some health indicators, such as the cardiac rhythm and blood pressure.

*Sharing education outside the school, city, country*

Another known opportunity provided to students by means of existing information technologies is the permanent and distance education using the Internet, which is not limited by the time frames of school schedule and by localization of the lessons inside classroom. Groups of students can also be extended beyond the school, city or country. Browsing the Web makes it possible to conduct lessons outdoors or in a museum. Thus, students with a smartphone create audio or video reports and submit them for the teacher review. The participation in the proposed project adds an investigation character to the boring studies. Under the teacher supervision the students can carry out various experiments that are performed and tested by sensors. Working with this software the students are forced to study the physical laws in practice.

*Automated monitoring of health and educational activity of students*

Specially configured smartphones enable to trace remotely the health status of students while they perform their individual works or during trips in the campaigns, or during sports trainings. These data, as well as results of the works or sports trainings have to be sent through Internet to the "school cloud". That cloud processes the data and, in the case of any abnormalities, sends back hints to student or alarm signals to teacher or parents. It could also give recommendations to students, help teachers to evaluate their work, parents to monitor the individual work of children and school doctors to help children timely detect possible health problems.

**What classes can use it**

We believe that measurements using of smartphones could improve processes of studying of the following subjects.

*Mathematics* – Introduction to the laws of geometry, trigonometry and by means of measuring distances and angles on the terrain.

*Physics* – Visual observation of some phenomena which are described the laws of physics and experimental verification of relevant mathematical formulas. Observations and measurement of invisible radiation and electromagnetic forces.

*Life science* – Visual observations, photos and videos fixation of environmental changes influenced by human activity.

*Geography* – Practical development of the geo-information-system (GIS) by mapping the terrain measurements of environmental parameters.

*Biology* – Discover and monitor the cardiac rhythm and processes that regulate the blood circulation in humans.

*Physical culture* – Accounting for effectiveness of training, tracking the pulse and blood pressure depending on the load.

The process of learning becomes a part of the game and at the same time creates a knowledge base that can be used anywhere – in the school and in local government, and environmental monitoring and scientific researches related to global change monitoring of environmental parameters.

**Conclusions**

The proposed new technology of global monitoring based on smart-phones and internet "clouds" could be developed in to the scientific, medical, or educational direction. The underlying information technology is just the same in all the cases. It includes the "cloud" calculations as intermediate stage and it opens a new wide possibilities of IT implementation. Possible inaccuracy in the data obtained by the simple sensors used in smartphones could be compensated by good statistical proving of the measured values.